\documentclass[twocolumn]{IEEEtran}

\usepackage{psfig}
\usepackage{epsfig}
\newcommand{\be}{\begin{equation}}
\newcommand{\ee}{\end{equation}}

\def\BibTeX{{\rm B\kern-.05em{\sc i\kern-.025em b}\kern-.08em
    T\kern-.1667em\lower.7ex\hbox{E}\kern-.125emX}}

\begin{document}

\title{How Traders enter the Market through the Book} 

\author{Lorenzo Matassini and Fabio Franci}

\maketitle

\begin{abstract}
We propose a simulation of a trading activity based on the implementation of the book, the list where all the orders
to buy and to sell a stock are recorded. Traders who do not own shares want to buy and agents who own commodities
want to sell them: The key factor here is of course the price. Everyone wants to be able to reach a positive spread
in order to make money. The main ingredients of our model are the price formation when entering an order and the
broadcasting of advertisements coming from a global view of the book. The time series of the market price, defined
everytime that a matching of two complementary orders occur, shows the well-known correlated volatility and fat tails
of returns phenomena, as observed in empirical data.
\end{abstract}

\begin{keywords}
Evolutionary Computation, Econophysics, Artificial Financial Market.
\end{keywords}

\section{Introduction}

\PARstart{T}{he} dynamics of stockmarket oscillations is not yet completely understood. It is anyway clear that market
price returns resemble the scaling laws characterizing physical systems dominated by the interaction of a large number of 
units. In \cite{haken} it is shown that, since the details of the circumstances governing the expectations and decisions 
of all the individuals are unknown to the modeler, the behavior of a large number of heterogeneous agents may best be 
formalized using a probabilistic setting, taking into account that the properties of macro variables are not necessarily 
identical to those of the corresponding micro variables \cite{ramsey}.

In our model we introduce only one kind of trader, since we want the features of the time serie of the market price
(of course a macro variable) to derive from the collective behavior of the agents (micro variable) competing for
the same goal and sharing the same resources. The well-known adage \emph{time is money} applies perfectly here: The
way to success lies in picking the right price at the right moment. The book is the most suitable structure to take
into account the joint effect of these two factors. The insertion of an order implies the storage of the desired price
and orders are sorted according to their arrival time.

Every trader has a limited amount of money and a given propension towards investment. The simulation starts with the
\emph{Initial Public Offer} ({\bf IPO}): All the shares belong to one agent, supposed to be the bank responsible of
the new emission. During this time, lasting until the bank does not own any share more, traders are encouraged to buy:
The buying list becomes longer and longer and agents get the shares at the IPO price, the price decided by the bank
for the emission. Some people may want to sell shares before the end of this phase, but the occurrence of such an event
is unlikely, since they ask a price bigger than the bank (this is the soul of speculation).

The bank cannot operate on the market when this transient is finished; who owns shares want to sell them for a
higher price and agents without shares want to buy them at a fair price. Of course the meaning of \emph{fairness} is
very different for buyers and sellers and this price formation represents one of the key aspect of the model.
Every trader, when willing to buy, has to identify the price according to past trends, opinion of the media and 
indications coming from competitors. The price coming out in the order is a weighted sum of the previous. The quality of
an investment has to take into account the involved time, too. That is why agents are also characterized by a threshold,
a kind of critical age of the share: They do not want to own it for a longer time than what is justified by the money
brought from the investment.

\section{Building the Book}

All the simulations presented here have been performed considering $N$ traders trading just $1$ stock with $M$ shares 
on a common market. A detailed description of the data structure for the \emph{type\_trader} is given in Appendix.
Here we comment only part of the field in that record. The algorithm is carried out in the following way: At each time 
step a trader is randomly selected and, according to the parameter \emph{investment\_inclination} the decision whether 
he/she will trade or not is taken. In this way we model the difference between intraday speculators and people who do 
not like to perform too many operations.

Through the parameter \emph{neighbours[]} we model the herding behavior, since agents share some information with 
acquaintances. To the same aim we have introduced \emph{memory[]}: The noise trader philosophy concerns with the analysis 
of past values of the market price looking for trends. These values are not constant, since not all the people have the 
same attitude to ask others for opinions and the amount of information they can consult about the stock varies from agent 
to agent, too.

Suppose now that the selected trader really wants to trade, namely he/she wants to sell (buy) some shares. In this case 
a new entry in the book list has to be created. Orders are stored according to the type (buy or sell), to the price and 
to the submission time. At every time step, after the trader's query, we also check the book state, looking for the 
occurrence of a transaction: It happens every time that the cheapest price among the sell list equals the most expensive 
offer in the buying list and this defines the market price in that moment.

When formulating the price $p$ for the order, every trader performs a weighted average over the following values:

\begin{itemize}
\item $p\_acq$, the target price averaged over all the acquanintances of the given agent.
\item $p\_bro$, the target price broadcasted by the media.
\item $p\_tre$, the expected price due to the actual trend, formulated taking into account the trend over the
               past \emph{MEM} prices.
\end{itemize}

The value of $p\_bro$ is generated considering the last market price $p\_las$ and comparing the current ratio between the 
number of buyers and the number of sellers in the book ($b\_unb$ stands for book unbalance) with the parameter $B$ 
(unbalance of the book), in the following way:
$p\_bro = p\_las (1 + g_{max})$ if $b\_unb > B$,
$p\_bro = p\_bro$ if $\frac{1}{B} < b\_unb < B$,
$p\_bro = p\_las (1-g_{max})$ if $ b\_unb < \frac{1}{B}$.

\begin{table}
\begin{center}
\begin{tabular}
{|p{0.8cm}|c|c|p{0.8cm}|p{0.8cm}|c|c|p{0.8cm}|}
\hline
\multicolumn{4}{|c|}{BUYING ORDERS} & \multicolumn{4}{|c|}{SELLING ORDERS}\\
\hline\hline
time & trad. & sh. & price & price & sh. & trad. & time\\
\hline
11012 & 120 & 5 & 10912 & 10914 & 6 & 652 & 10802\\
\hline
10951 & 1 & 25 & 10910 & 10914 & 16 & 12 & 11115\\
\hline
11435 & 233 & 7 & 10909 & 10916 & 8 & 431 & 11613\\
\hline
10702 & 890 & 3 & 10906 & 10917 & 1 & 801 & 10211\\
\hline
11647 & 507 & 4 & 10905 & 10920 & 3 & 212 & 11765\\
\hline
\end{tabular}
\caption{\label{tab:esempiobook} Example of the first five levels of the book. No transaction can take place because the 
highest buying price is smaller than the cheapest selling order. Entries are ordered according to price and occurrence 
time.}
\end{center}
\end{table}

To get the fair price with which entering the market, every trader subtracts a given quantity to $p$, according to his/her 
\emph{expected\_gain}. 
When querying an agents who is still waiting for the execution of a submitted order, he/she can also decide to change some
parameter of it, considering the elapsed time and all the other information already discussed in this section. 

In Tab.\ref{tab:esempiobook} we give an example of the first five levels of the book. We can see that orders are 
divided according to the type, namely buy or sell, and ordered on the basis of price (decreasing order for the buy side,
increasing order for the sell side), time of arrival. Therefore the first line of the book contains the highest buying 
price and the cheapest selling price. Since the second value is bigger than the first, there is no chance for any 
transaction to occur. For more details we refer the reader to the Appendix, where the pseudo\_code for the algorithm is 
presented.

\section{Getting the Market Price Time Series}

\begin{figure}
\centerline{\psfig{file=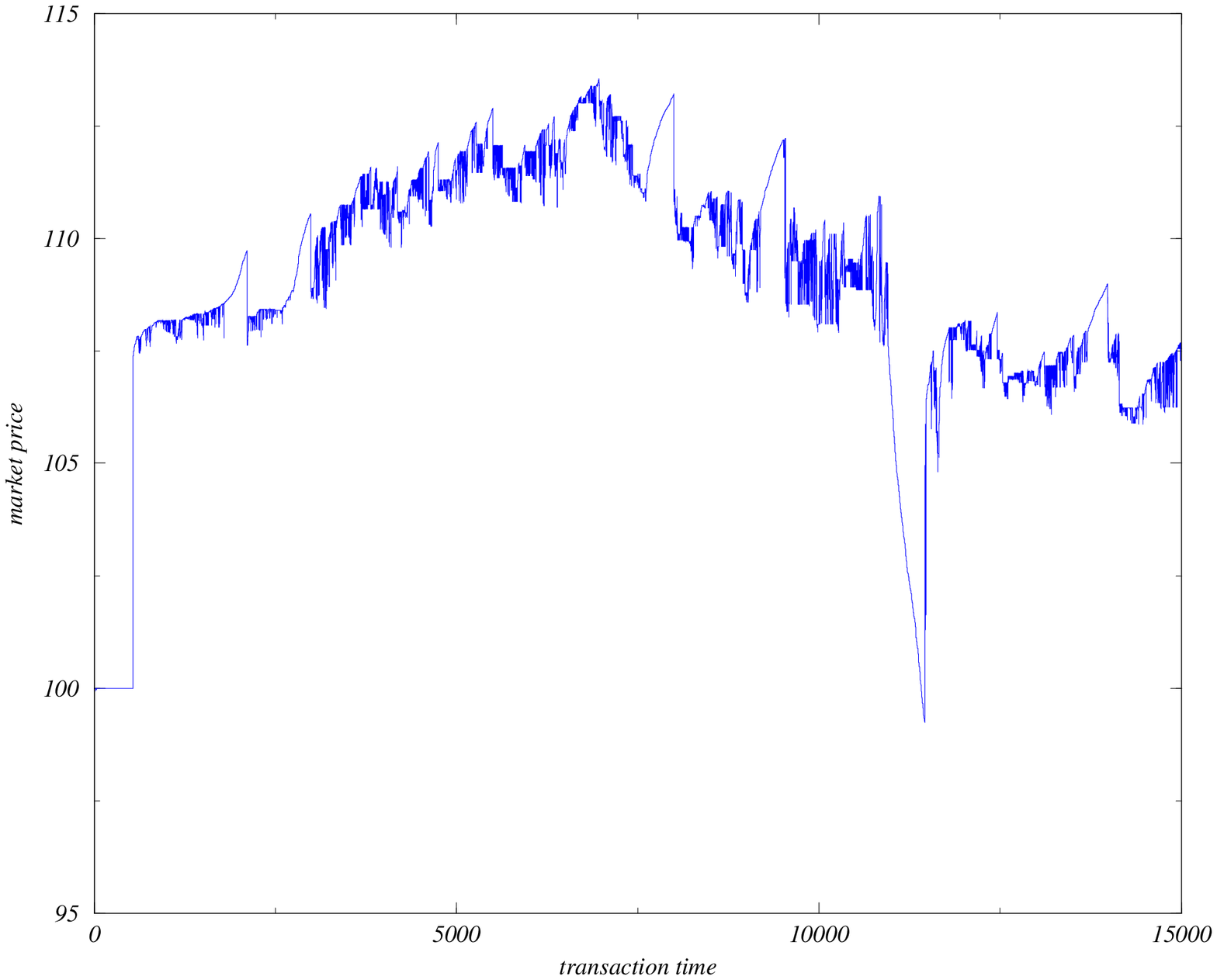,width=7cm,angle=270}}
\caption[]{\small\label{fig:serie}
Typical snapshot from a simulation run. Development with transaction time of the market price.}
\end{figure}

The result of a typical simulation run is shown in Fig.\ref{fig:serie}: We can clearly see the IPO phase lasting few 
hundreds of ticks, where the price remains constant because the bank offers the shares to the traders at a suggested 
value. After this transient, the pressure made from people without shares forces the price to raise. Letting the time 
evolve, one runs into a tranquil period, with a slowly oscillating price, since owners do not want to sell because they 
hope to get more money if the wait a little bit more and agents without shares want to buy at a lower price.
This dynamical equilibrium is unstable and leads to a small burst whose effect is that the price increases to a more 
suitable value for owners. A rally is usually followed by a crash (a kind of settlement),  as reported in \cite{platt} 
under the name of \emph{on-off intermittency}, an aperiodic switching between static, or laminar, behavior and chaotic 
bursts of oscillations. On the other hand a rally usually follows a crash, what is happening in the last part of
Fig.\ref{fig:serie}.

\begin{figure}
\centerline{\psfig{file=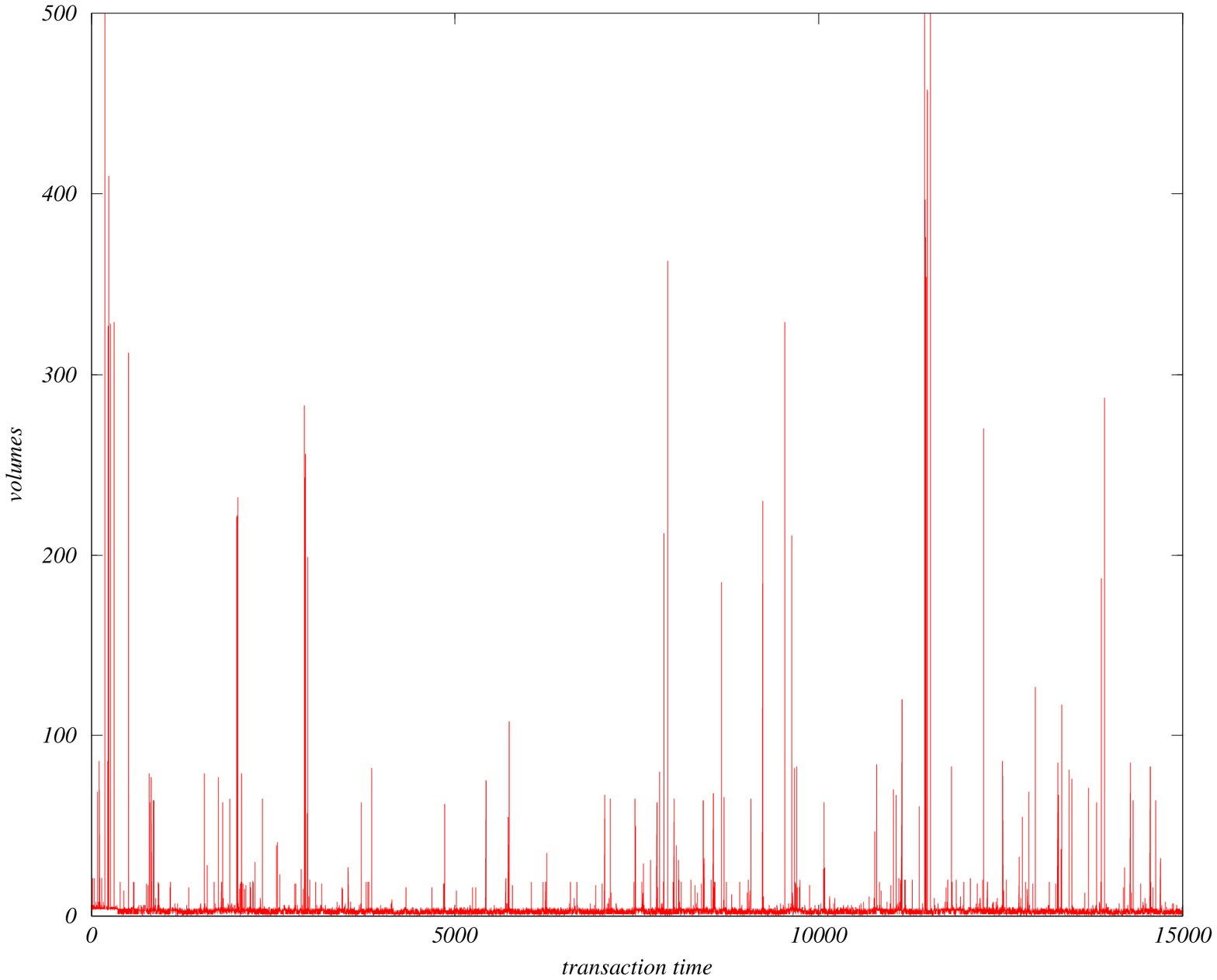,width=7cm,angle=270}}
\caption[]{\small\label{fig:volumi}
Typical snapshot from a simulation run. Development with transaction time of the total exchanged shares (volumes).}
\end{figure}

In Fig.\ref{fig:volumi} we have plotted the corresponding volumes of Fig.\ref{fig:serie}. The correlation between the
total amount of exchanged shares and the price variations is clear. It becomes particularly evident in the last part
of the graph, where the big crash occurs: A lot of people would like to sell shares, but they can do that only subject
to a cheapest request; as the price starts to decrease some agents may sell as a consequence of their stop-loss
constraints and this produces a further diminution of the market price, such that many of the buying orders can be
performed. These new owners do not sell immedietely the shares and this helps to escape from the \emph{panic selling
phase}: The crash is finished and immediately followed by a rally period.

As already mentioned, universal characteristics exhibited by financial prices comprise a distribution with fat tails
(events with a distance bigger than $3\sigma$ from the average return are not so uncommon as a Gaussian distribution
would forecast) and a correlation in the volatility (alternation between tranquil and tumultuous periods). 
In Tab.\ref{tab:hurst} we report on the presence of a strong persistence in the volatility. This is done 
estimating the self-similarity parameter H \cite{peng} for raw and absolute returns (being the latters a measure 
of volatility). The behavior of our simulation is in total agreement with empirical results, as proved by H going
from 0.52 to 0.85. A random walk shows no significant increase of H when switching from raw returns to absolutes ones.

\begin{table}
\begin{center}
\begin{tabular}{|l|c|c|}
\hline
& Raw returns & Absolute returns\\
\hline\hline
financial & 0.49 & 0.84\\
\hline
simulation & 0.52 & 0.85\\
\hline
random & 0.50 & 0.53\\
\hline
\end{tabular}
\caption{\label{tab:hurst} Hurst exponent of raw and absolute returns for different time series.}
\end{center}
\end{table}

\begin{figure}
\centerline{\psfig{file=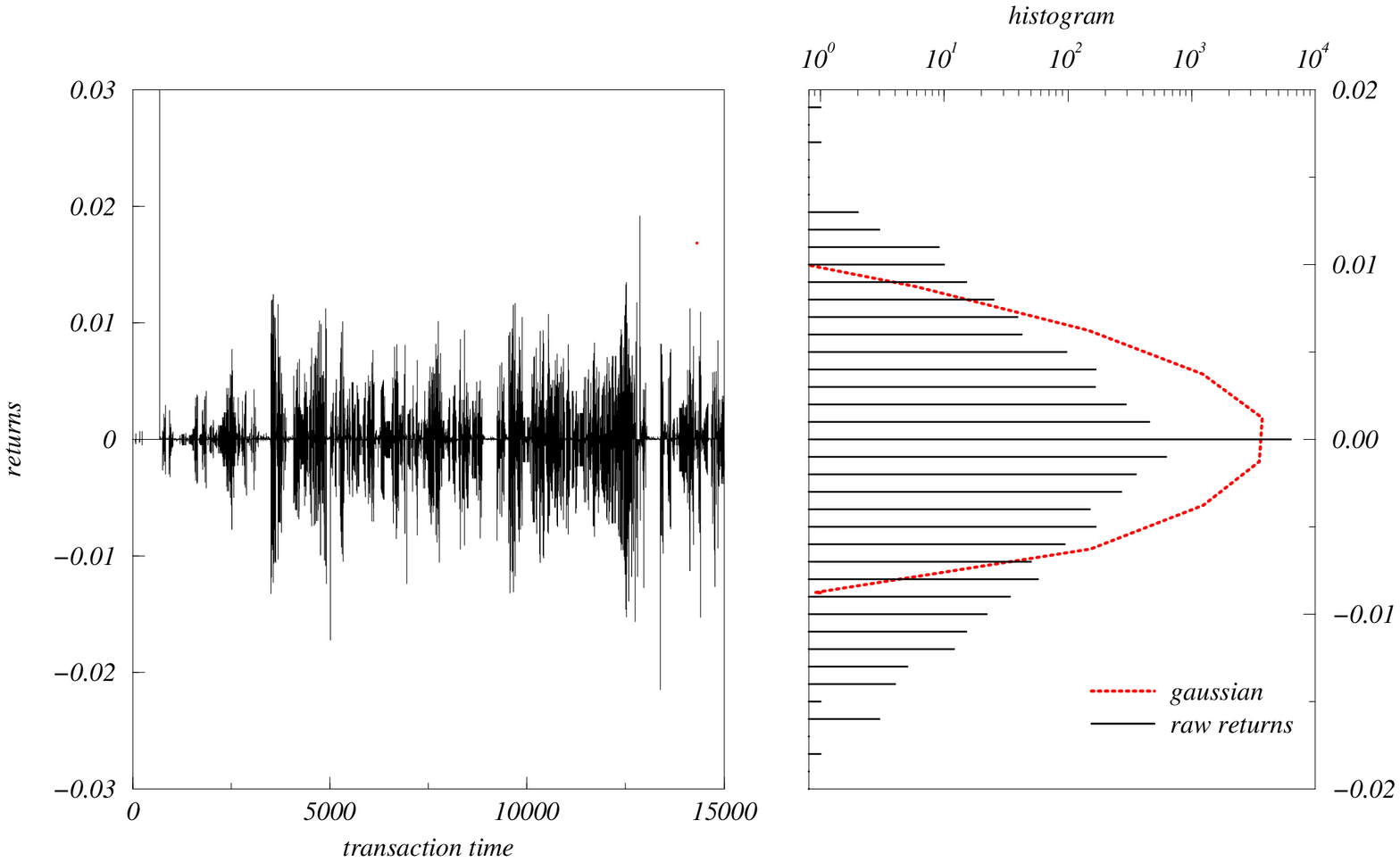,width=6cm,angle=270}}
\caption[]{\small\label{fig:fattails}
Left panel: Price returns. Right panel: Returns distribution. The comparison with a best-fitted normal distribution 
(dotted line) reveals the presence of \emph{fat tails}.}
\end{figure}

Fat tails are detectable through the probability density function (PDF) of the returns. In Fig.\ref{fig:fattails} we
show the PDF of our simulated time series together with a gaussian distribution of returns having the same standard
deviation. Extreme events happen with a higher frequency in the simulation, giving raise to the typically observed 
alternation between tranquil period, rally days and crashes.

We want now to discuss a problem found in evaluating the richness of traders. Every agent can possess cash money
and shares. The value of the former is obvious, but concerning the latter some problem arises: If one has 100 shares,
which is the equivalent amount of richness, the actual market value or maybe the original buying price? If the number
of shares is small compared to the total volumes, then one can be almost surely able to sell them at the market price,
but what happens if too many people want to do the same simultaneously? This problem is similar to the Taylor series,
whose validity is limited to a small neighbourood of the central point. So, if we want to evaluate the richness of
all the traders at a given time we have to cope with the following problem: Which value should we give to each share?

To illustrate it in more details, we can consider what happens if we evaluate the richness of a trader as given
by the sum of the cash and the invested money, i.e. associating the original buying price to the shares. Considering 
the invested money as real money we suppose (without any reason) that the trader is able to sell the shares exactly 
at the buying price. As a consequence of this assumption, the virtual richness increases and decreases together with 
the market price, with a small delay. We call therefore virtual this richness.
But how can we conciliate the fact that the market is a closed one with such a variability of the richness?
The reason is the assumption about the value of the shares and we can easily prove that the virtual richness comes back
to the starting value if and only if the market price comes back (and remains constant a sufficiently long
period of time) to the IPO value in order to let all the people perform transactions at this market price.
But evaluating the richness of a trader as the sum of cash plus number of shares time the IPO price we neglet all the 
dynamics of the market price. Furthermore when we decide to stop the simulation we impose an artificial behavior to our
simulated market without any equivalence with the real life.

To solve this evaluating problem we should provide a mechanism that allow all the traders to sell the shares to a special
\emph{buy-back bank} exactly at the IPO price. Since the agents do not know when the end of the simulation will occur,
they cannot take into account this buy-back procedure and if the market price is above (below) the IPO value, they will
loose (gain) money without any reason. We will address this problem in a future work, for the moment we just wanted to
point out the difficulties arising when trying to evaluate the richness in a financial market context.

\section{Conclusions}

We have simulated a stockmarket through the costruction of the book for buying and selling orders. Our model is able to 
reproduce the two main characteristics of empirical data, namely correlated volatility and fat tails of the PDF of returns,
with only one type of traders in competition with similar strategies, similar resources and similar whishes.
They do not want to wait too much before they can have back the invested money plus a significant gain, they state clearly
through the inserted orders which is the minimum gain they would like to get and up to which losses they can continue to
play. Furthermore they do not invest all the money they have at disposal, since they want to use it in the most appropriate
and less risky way.
These conditions have been implemented thanks to the strategy of price formation and the broadcasting of advertisements,
mechanisms with which we provide both a local and a global coupling, responsible of the good agreement with empirical
data. The key aspect here is the interplay between time and money, namely between the price formation and the insertion
of the order.

\appendix{Pseudo-code}

\subsection{Parameters}
\begin{verbatim}
#define NUM_TRADER 1000 
#define THRESHOLD 10000
#define NUM_MAX_NEIGHBOURS 10
#define NUM_MIN_NEIGHBOURS 1
#define GAIN_MIN 0.05
#define GAIN_MAX 0.15
#define LOSS_MIN 0.03
#define LOSS_MAX 0.10
#define INVESTMENT_INCLINATION_MIN 0.2 
#define INVESTMENT_INCLINATION_MAX 0.5
#define NUM_STOCKS 5000 
#define IPO_PRICE 10000   
#define MINIMAL_RESOURCES 100000       
#define TOTAL_ZIPF_RESOURCES 100000000   
#define SIMULATION_TIME 1000000  
#define MAX_MEMORY_LENGTH 50    
#define BOOK_UNBALANCE 2
#define ZIPF1 0.2 
#define ZIPF2 0.8
\end{verbatim}

\subsection{Data structures and variables}
\begin{verbatim}
struct type_news
{
  unsigned int target_price;
  char rating; /*1=buy 2=neutral 3=sell */
} news;

struct type_order
{
  int stocks;
  unsigned int price;
  unsigned int id;
  unsigned long tick;
  struct type_order *next;
} ;

struct type_order *buyers_book=NULL; 
struct type_order *sellers_book=NULL; 

struct type_trader
{
  unsigned long initial_resources;  
  unsigned long total_resources; 
  unsigned long invested_resources; 
  double p_influence; 
  unsigned int stocks; 
  unsigned int neighbours[NUM_MAX_NEIGHBOURS];
  unsigned int num_neighbours; 
  unsigned int memory[MAX_MEMORY_LENGTH];
  unsigned int memory_max; 
  double personal_threshold;
  double investment_inclination;
  double liquidity_inclination; 
  double expected_gain;  
  double maximum_loss; 
  unsigned int target_price;
  unsigned int stop_loss;
  char order_state; 
     
} trader[NUM_TRADER];
\end{verbatim}

\subsection{Data flow}

\subsubsection{Initialization of the traders}
All the features of the traders are initialized randomly among the limits
difined by the parameters.
The resources are distributed according to the ZIPF's Law. The meaning of the
ZIPF's parameters is that the \verb+ZIPF1+ of traders owns the \verb+ZIPF2+ of
the total resources.
Beside this, every trader is given \verb+MINIMAL_RESOURCES+ so that his/her 
\verb+initial_resources+ is not zero.

\subsubsection{IPO-Phase (transient)}
The \emph{bank} responsible for the IPO is selected randomly among the traders:
\begin{verbatim}
IPO_TRADER=rand()*(NUMERO_TRADER-1)/RAND_MAX;
trader[IPO_TRADER].stocks=NUM_STOCKS;
saved_res=trader[IPO_TRADER].total_resources;
trader[IPO_TRADER].total_resources=0;
trader[IPO_TRADER].invested_resources=
NUM_STOCKS*IPO_PRICE;
sell_order(0,IPO_TRADER,NUM_STOCKS,IPO_PRICE);
\end{verbatim}

News are broadcasted to convince traders to buy some shares:
\begin{verbatim}
news.target_price=IPO_PRICE*1.1;
news.rating=1;
\end{verbatim}

Then the market evolution starts until the IPO-Phase is over:
\begin{verbatim}
while (trader[IPO_TRADER].stocks>0)
{
  ...
\end{verbatim}  
  (This part is explained in the \emph{Regime} section.) 
\begin{verbatim}  
}
\end{verbatim}

At the end of the IPO-Phase the \verb+news+ are restored to
more neutral positions:
\begin{verbatim}
news.target_price=IPO_PRICE;
news.rating=2;
\end{verbatim}

and the initial condition of the ipo\_trader is restored
\begin{verbatim}
trader[IPO_TRADER].total_resources=saved_res;
trader[IPO_TRADER].initial_resources=saved_res;
trader[IPO_TRADER].invested_resources=0;
\end{verbatim}

\subsubsection{Regime}
\begin{verbatim}
while (time<SIMULATION_TIME)
{
  time++;
  position=(double)rand()*(NUM_TRADER-1)/RAND_MAX;
  operativity(position);
  spread_control(0);
}
\end{verbatim}

Here we describe the procedure \verb+operativity+:

\begin{verbatim}
operativity(unsigned int pos)
{
  if (trader[pos].order_state==0)
  { /*no pending orders*/
    if (trader[pos].stocks==0)
    { /*no shares owned*/
      prob=(double)rand()/RAND_MAX;
      if (prob<trader[pos].investment_inclination)
      {
        price=price_formation(pos);
        trader[pos].target_price=price*
         (1+trader[pos].expected_gain);
        trader[pos].stop_loss=price*
         (1-trader[pos].maximum_loss);
        stocks=(trader[pos].liquidity_inclination*
         trader[pos].total_resources)/price;
        buy_order(time,pos,stocks,price);
      }  
    }
    else
    { /*shares owned*/
      if (last_price<trader[pos].stop_loss)
      { /*stop loss application*/
        if (news.rating==3)
        {
          market_sell(pos);
        }
        else if (news.rating==1)
        {
          if (trader[pos].total_resources >
            trader[pos].liquidity_inclination*
             (trader[pos].total_resources+
              trader[pos].invested_resources))
          {
            average_buy(pos);
          }
        }
        else if (news.rating==2)
        {
          if (trader[pos].total_resources >
              trader[pos].liquidity_inclination*
              (trader[pos].total_resources+
               trader[pos].invested_resources))
          {
            average_buy(pos);
          }
          else
          {
            market_sell(pos);
          }
        }
      }
      else
      { /*insert sell order*/
        sell_order(time,pos,
         trader[pos].stocks,
         trader[pos].target_price);
      }
    }
  }
  else
  { /*pending order*/
    p1=search(sellers_book);
    if (p1)
    { /*pending selling order*/
      if (last_price<trader[pos].stop_loss)
      {
        remove_sell_order(p1);
      operativity(pos);
      }
      else if (time-p1->tick)>
      (THRESHOLD/personal_threshold) 
      {
        price=price_formation(pos);
        price*=(1+trader[pos].expected_gain);
        if (price<trader[pos].target_price)
        {
          remove_sell_order(p1);
          sell_order(time,pos,
          trader[pos].stocks,price);
        }
      }
    }
    else
    {
      p1=search(buyers_book);
      if (p1)
      { /*pending buying order*/
        if (time-p1->tick)>
        (THRESHOLD/personal_threshold)
        {
          expense=p1->stocks*p1->price;
          price=p1->price;
          remove_buy_order(p1);
          prob=(double)rand()/RAND_MAX;
          if (prob<
           trader[pos].investment_inclination)
          {
            market_buy(pos,expense);
            trader[pos].target_price=
             (trader[pos].invested_resources/
             trader[pos].stocks)*
             (1+trader[pos].expected_gain);
            trader[pos].stop_loss=
             (trader[pos].invested_resources/
             trader[pos].stocks)*
             (1-trader[pos].maximum_loss);
           }
           else
           {
              price=(1+trader[pos].expected_
			                           gain/2);
              stocks=expense/price;
              buy_order(time,pos,stocks,price);
           }
        }
      }  
      else
      {
        printf("Error! 
          No order in the books!\n");
        exit(1);
      }
    }
  }
}
\end{verbatim}

\begin{verbatim}
unsigned int price_formation (unsigned int pos)
{
  p_acq=0;
  for (ii=0;ii<trader[pos].num_neighbours;ii++)
  {
    if (trader[trader[pos].neighbours[ii]]
       .target_price)
    {  
      counter++;
      p_acq+=trader[trader[pos].neighbours[ii]]
        .target_price;
    }
  }
  p_acq/=counter;
  p_acq*=(1-trader[pos].expected_gain);
  counter=0;
  p_tre=0;
  for (ii=0;ii<trader[pos].memory_max;ii++)
  {
    if (trader[pos].memory[ii])
    {
      p_tre+=trader[pos].memory[ii];
      counter++;
    }
  }
  p_tre/=counter;
  p_bro=news.target_price*
    (1-trader[pos].expected_gain);
  
  price=a1*p_acq+a2*p_tre+a3*p_bro;
  return(price);
}
\end{verbatim}

The procedures:
\begin{verbatim}
buy_order();
sell_order();
market_buy();
market_sell();
average_buy();
remove_buy_order();
remove_sell_order();
\end{verbatim}
are used to insert orders in the book or remove them and are not reported
because simply realized with list operations.

The procedure \verb+spread_control()+ has the purpose to check the book entries
and compute the operations following a transaction, that is removing the entry
in the book, updating the resources and the stocks owned by the trader, storing
the last price and the volumes.

The procedure \verb+search()+ is used to find an order in the books.

News are generated according to the following rules:
\begin{verbatim}
evaluate_book();
if (buyers/sellers) > BOOK_UNBALANCE
{
  news.rating=1;
  news.target_price=
    last_price*(1+GAIN_MAX);
}
else if (sellers/buyers) > BOOK_UNBALANCE
{
  news.rating=3;
  news.target_price=
    last_price*(1-GAIN_MAX);
}
else
{
  news.rating=2;
}
\end{verbatim}

The procedure \verb+evaluate_book()+ has the purpose to scan the book and return the buyers' pressure and the
sellers' pressure, according to the number of orders and the number of shares involved.

\nocite{*}

\bibliographystyle{IEEE}

\begin{thebibliography}{1}

\bibitem {haken} Haken, H. Synergetics: An introduction. \emph{Springer, Berlin} (1983).

\bibitem {ramsey} Ramsey, J.B. On the existence of macro variables and macro relationships. \emph{Journal of Economic
                  Behavior and Organization} {\bf 30}, 275-299 (1996).

\bibitem {platt} Platt, N., Spiegel, E.A. \& Tresser, C. On-off intermittency: A mechanism for bursting.
                 \emph{Physical Review Letters} {\bf 70}, 279-282 (1993).
				 
\bibitem {peng} Peng, C.K., Buldyrev, S., Havlin, S., Simons, M., Stanley, H.E. \& Goldberger, A.L. Mosaic
                organization of DNA nucleotides. \emph{Physical Review E} {\bf 49}, 1685-1689 (1994).	

\end{thebibliography}

\end{document}